\documentclass[aps,pra,preprint,superscriptaddress,longbibliography]{revtex4-1}

\usepackage{amsmath}
\usepackage{amsthm}
\usepackage{amsfonts}
\usepackage{amssymb}
\usepackage{xcolor,graphicx}
\usepackage{tikz}
\usepackage{color}
\usepackage[pdftex]{hyperref}
\usepackage{longtable}
\usepackage{array}
\usepackage{bigints}
\usepackage{dsfont}
\begin{document}
\bibliographystyle{apsrev4-1}

\setkeys{Gin}{draft=false} 

\title{Generalized vectorial ray tracing for optical analysis and design in freeform gradient-index media}

\author{V. Ch\'{a}vez-Islas}
\email{vrani.chavez@inaoep.mx}
\affiliation{Instituto Nacional de Astrof\'isica, \'Optica y Electr\'onica, Santa María Tonantzintla, Puebla, 72840, Mexico}

\author{E. Esp\'{i}ndola-Ramos}
\email{ernestoespindola@inaoep.mx}
\affiliation{Instituto Nacional de Astrof\'isica, \'Optica y Electr\'onica, Santa María Tonantzintla, Puebla, 72840, Mexico}

\author{F. H. Maldonado-Villamizar}
\email{fmaldonado@inaoep.mx}
\affiliation{SECIHTI - Instituto Nacional de Astrof\'isica, \'Optica y Electr\'onica, Santa María Tonantzintla, Puebla, 72840, Mexico}

\author{S. Ch\'{a}vez-Cerda}
\email{sabino@inaoep.mx}
\affiliation{Instituto Nacional de Astrof\'isica, \'Optica y Electr\'onica, Santa María Tonantzintla, Puebla, 72840, Mexico}

\author{J. E. G\'{o}mez-Correa}
\email{jgomez@inaoep.mx}
\affiliation{Instituto Nacional de Astrof\'isica, \'Optica y Electr\'onica, Santa María Tonantzintla, Puebla, 72840, Mexico}



\begin{abstract}
Gradient-index (GRIN) media with freeform refractive-index distributions require ray tracing approaches capable of accurately modeling light propagation in complex three-dimensional environments. In this work, we develop a fully three-dimensional ray tracing method for arbitrary freeform GRIN media based on the local application of a set of vectorial relations of geometrical optics. At its core, the method constructs ray trajectories through successive local refractions using the vectorial form of Snell's law, while employing constant optical path length steps that naturally adapt the geometric step size within the medium. Unlike conventional continuous GRIN propagation approaches, the proposed method naturally handles refraction at the boundary of the medium as well as total internal reflection, enabling the treatment of both continuous media and discretized refractive-index distributions composed of isoindicial surfaces. The method is simple to implement and computationally efficient. Validation against an analytical solution demonstrates high accuracy, while applications to freeform GRIN configurations illustrate its capability not only for propagation analysis, but also for the direct design of complex tailor-made optical media.
\end{abstract}
\maketitle




\section{Introduction}

In 1972, O. N. Stavroudis, in his book \textit{The Optics of Rays, Wavefronts, and Caustics}, characterized geometrical optics as “seventeenth and eighteenth century physics camouflaged behind nineteenth and twentieth century mathematics” \cite{Stavroudis1972}. At first glance, in light of modern developments such as quantum optics, nanophotonics, and relativity, one might be tempted to regard geometrical optics as an outdated framework with limited relevance in the twenty-first century.

However, this view is misleading. In quantum optics, a large class of experiments is implemented using optical systems—such as lenses, interferometers, and imaging setups—whose design and analysis rely fundamentally on the framework of classical optics, including geometrical optics \cite{Knight2005,MandelWolf1995}. In nanophotonics, although many phenomena occur in regimes where wave effects are essential, the high-frequency (short-wavelength) limit leads naturally to ray-based descriptions through the eikonal approximation, and energy transport can often be interpreted in terms of effective rays or trajectories governed by the local structure of the field \cite{NovotnyHecht2025,BornandWolf1999}. Likewise, in general relativity, the propagation of light in curved spacetime is described by null geodesics, which arise as the geometrical-optics limit of Maxwell’s equations in a curved metric, thereby providing a natural extension of ray optics to non-Euclidean geometries \cite{Synge1960,Revisiting2025}.

Geometrical optics naturally emerges as the short-wavelength limit of wave optics, providing an asymptotic framework that remains valid whenever the characteristic dimensions of the system are large compared to the wavelength \cite{BornandWolf1999,ContemporaryGhatak}. This interpretation is supported by classical works such as those of Luneburg \cite{Luneburg}, and later Kline and Kay \cite{Kline1965}, where geometrical optics is understood as an approximation to the electromagnetic theory of light. In this sense, geometrical optics is not a separate or outdated theory, but rather a limiting case of a more general physical framework.

As emphasized by Stavroudis, regardless of our acceptance of its underlying assumptions, geometrical optics retains a unique position in modern technology \cite{Stavroudis1972}. It remains the only practical framework through which the global properties of an optical system can be directly related to its design parameters. In this sense, geometrical optics is not merely an approximation, but the natural language of optical design, enabling intuitive control over light propagation and system performance.

This perspective becomes particularly relevant in the context of gradient-index (GRIN) media, where the refractive index varies continuously in space and the control of ray trajectories is intrinsically linked to the spatial distribution of the index \cite{GomezLibro,Marchand,Lakshminarayanan}. As advances in fabrication and modeling enable increasingly complex freeform GRIN structures \cite{Lippman:21}, the analysis of light propagation in arbitrary refractive-index distributions becomes increasingly important.

Ray tracing methods in GRIN media can be broadly classified into analytical, quasi-analytical, and numerical. Analytical methods provide exact solutions to the ray equations; however, they are restricted to specific refractive index distributions for which closed-form solutions exist \cite{Marchand:72,Streifer:71,Paxton:71, Buchdahl:73}. Quasi-analytical approaches, including those based on invariants derived from the Euler--Lagrange equations applied to the optical Lagrangian, extend this framework but still rely on particular symmetries or approximations that limit their general applicability \cite{Marchand:70, Moore:75, Marchand:88,Gomez-Correa:21,Gomez-Correa:22,Liu:19}. Numerical methods, on the other hand, offer greater flexibility and can handle arbitrary refractive-index distributions. In many cases, these approaches describe ray propagation through the numerical integration of differential equations derived from geometrical optics principles \cite{Montagnino:68,Rawson:70,Yu:24}.

From a physical standpoint, throughout nearly its entire propagation in a GRIN medium, light propagation consists of a continuous sequence of local refraction events. This viewpoint suggests that ray trajectories may be constructed through successive local applications of Snell’s law, provided that the propagation step is sufficiently small \cite{Chavez-Islas:26,Chavez-Islas:25}. Such a formulation describes ray evolution directly in terms of refractive interactions, providing an intuitive picture of propagation within the medium.

In this paper, motivated by this interpretation, we introduce Vector Ray Tracing (VRT), a framework that models ray propagation through the successive application of a set of vectorial relations of geometrical optics. Combined with a propagation scheme based on constant optical path length (OPL) increments, the proposed formulation provides a unified treatment of light propagation in homogeneous media, discretized refractive-index distributions, and continuous GRIN media. As a result, phenomena such as interface refraction, total internal reflection, and gradient-induced ray deviations can be described within the same framework. Because the method is formulated directly in terms of vectorial refraction, it remains straightforward to implement for arbitrary three-dimensional freeform GRIN media. Beyond propagation analysis, the proposed framework is employed as a practical tool for optical design.

\section{ Vectorial Ray Tracing (VRT)}

The law of refraction, widely attributed to the empirical work of Willebrord Snell before 1621, was indeed first formally described by the Persian mathematician Ibn Sahl in the 10th century \cite{Rashed1990API} through geometric insights into burning lenses. Although Descartes is often credited with the first attempt to derive the law \cite{Joyce:76}, his awareness of Snell's work remains a subject of controversy. Beyond this historical evolution, Snell's law remains one of the most fundamental principles in optics 
\cite{Stavroudis1972,Herzberger:66}. Together with the law of reflection, they constitute fundamental constraints for any physically consistent description of light propagation. Any theoretical framework describing light must be consistent with these laws; otherwise, it cannot be considered a physically valid description of light propagation and should therefore be abandoned \cite{Stavroudis1972}.

In its classical form, Snell’s law is typically expressed in terms of scalar angles measured with respect to the interface normal \cite{BornandWolf1999,ContemporaryGhatak} \begin{equation}
n_{\mathrm{inc}} \sin\theta_{\mathrm{inc}} = n_{\mathrm{ref}} \sin\theta_{\mathrm{ref}},
\end{equation}
where $n_{\mathrm{inc}}$ and $n_{\mathrm{ref}}$ are the refractive indices of the incident and refractive media, respectively, while $\theta_{\mathrm{inc}}$ and $\theta_{\mathrm{ref}}$ denote the angles formed by the incident and refracted rays with respect to the interface normal. While this formulation is sufficient for simple two-dimensional configurations, it becomes inadequate for describing ray propagation in fully three-dimensional and spatially varying media. In such cases, a vectorial formulation is required to account for the complete directional behavior of light \cite{Stavroudis1972,Bhattacharjee_2005}.

The relevance of Snell’s law extends beyond its role as a local constraint; it can be generalized to gradient-index media by considering each iso-indicial surface as the interface between two adjacent layers of differential thickness. Consequently, within our VRT framework, it constitutes the fundamental mechanism through which ray trajectories are constructed. Rather than obtaining ray paths from the integration of differential equations, propagation is explicitly generated as a sequence of refraction events, where the ray direction is updated iteratively at each step. In this way, Snell’s law is transformed into a step-by-step propagation rule.

\begin{figure}[t]
\centering
\includegraphics[width=0.5\linewidth]{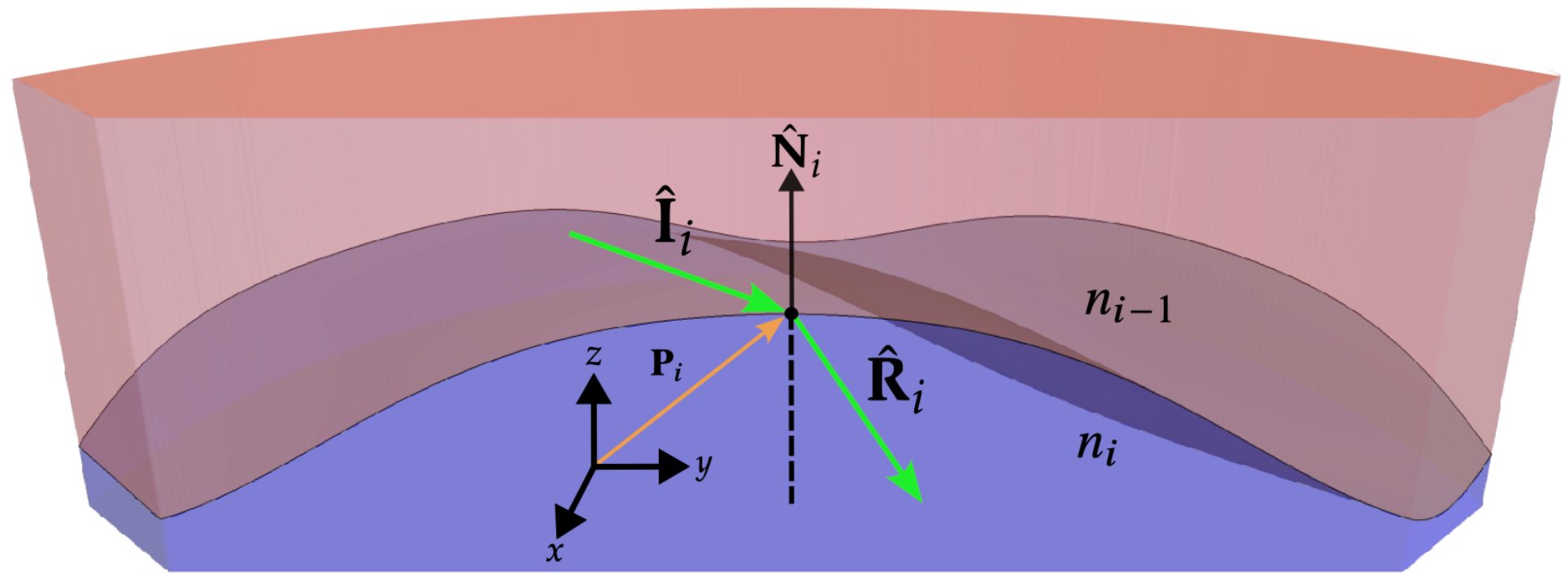}
\caption{Configuration of the incident vector $\hat{\mathbf{I}}_i$, refracted vector $\hat{\mathbf{R}}_i$ and normal vector $\hat{\mathbf{N}}_i$ at the refraction point $P_i$, located at the interface between two media with refractive indices $n_{i-1}$ and $n_i$. Successive ray segments are constrained to satisfy Snell’s law; consequently, the incident, refracted, and normal vectors are locally coplanar.}
\label{fig:1}
\end{figure}

Within this framework, refraction is naturally described in terms of the incident, refracted, and normal unit vectors associated with isoindicial surfaces, leading to the vectorial representation of Snell’s law. Accordingly, this law can be written as \cite{Stavroudis1972,Bhattacharjee_2005,stavroudis2006mathematics}
\begin{eqnarray}
\label{refraction_law}
    n_{i-1} \hat{{\bf N}}_{i}\times \hat{{\bf I}}_{i}=n_{i} \hat{{\bf N}}_i\times\hat{{\bf R}}_i,
\end{eqnarray}
here $i$ denotes the segment index along the discrete ray trajectory, $\hat{\mathbf{I}}_{i}$ is the direction of the incident ray propagating through the medium with refractive index $n_{i-1}$, $\hat{\mathbf{R}}_{i}$ is the direction of the refracted ray in the homogeneous medium of refractive index $n_i$, and $\hat{\mathbf{N}}_{i}$ is the unit normal to the interface separating the  homogeneous regions at point $\mathbf{P}_{i}$, where the ray is refracted; therefore, $\hat{\mathbf{N}}_i$, $\hat{\mathbf{I}}_i$, and $\hat{\mathbf{R}}_i$ are coplanar, as depicted in Fig.~\ref{fig:1}.

The VRT is then performed by advancing the ray iteratively from one point to the next. Once the refracted direction $\hat{\mathbf{R}}_{i}$ is determined, the ray is propagated according to
\begin{eqnarray}
\label{ray_evolution}
    \mathbf{P}_{i+1} =\mathbf{P}_i + \Delta_i \hat{\bf{R}}_i,
\end{eqnarray}
where the geometrical step length is defined as
\begin{eqnarray}
    \Delta_i = \frac{\Delta L_0}{n_i}
\end{eqnarray}
 and $\Delta L_{0}$ is the OPL increment that will remain constant on propagation. This choice ensures that the whole propagation remains consistent with the physical definition of OPL, naturally adapting the geometrical step size to local variations of the refractive index. In this way, the ray trajectory is constructed as a sequence of discrete steps, each governed by a refraction event. 
 
As mentioned above, we will consider that the medium is such that the separation between isoindical surfaces is  infinitesimal allowing the assumption that the medium layer in between is homogeneous. Consequently, if $n(x,y,z)$ is the refractive index distribution, then $n_{i-1} = n(\mathbf{P}_{i-1})$ denotes the values of the $n-1$ layer where the incident ray is traveling, while $n_i = n(\mathbf{P}_i)$ corresponds to the following layer where the local refraction occurs.

The local geometry of the GRIN medium at $\mathbf{P}_i$ is characterized by the gradient of the refractive index, which defines the unit normal as
\begin{equation}
\label{normal}
    \hat{\mathbf{N}}_i = \left. \frac{\nabla n}{\|\nabla n\|} \right|_{\mathbf{P}_{i}}.
\end{equation}
At the same point, the incident direction is obtained from the previous segment of the trajectory as
\begin{equation}
\hat{\mathbf{I}}_i = \frac{\mathbf{P}_i - \mathbf{P}_{i-1}}{\|\mathbf{P}_i - \mathbf{P}_{i-1}\|}.
\end{equation}

Once the incident direction $\hat{\mathbf{I}}_i$ and the unit normal $\hat{\mathbf{N}}_i$ are established, the direction of the refracted ray $\hat{\mathbf{R}}_i$ must be determined. Following Stavroudis, Eq.~(\ref{refraction_law}) can be rewritten as \cite{Stavroudis1972}
\begin{equation}
    \hat{{\bf N}}_i \times \left( \hat{{\bf R}}_i -\mu_i \hat{{\bf I}}_i  \right)=0,
\end{equation}
where $\mu_i =n_{i-1}/n_i$. Therefore, the resultant vector inside the parentheses must be collinear with the normal vector; that is to say, there must exist a scalar quantity $\gamma$ such that
\begin{equation}
    \hat{{\bf R}}_i -\mu_i \hat{{\bf I}}_i = \gamma_i \hat{{\bf N}}_i,
    \label{Eq:Seven}
\end{equation}
thus, the direction of the refracted ray is given by
\begin{eqnarray}
\label{refracted_ray}
    \hat{{\bf R}}_i  = \mu_i \hat{{\bf I}}_i +\gamma_i \hat{{\bf N}}_i,
\end{eqnarray}
where
\begin{equation}
\label{gamma}
    \gamma_i = -\mu_i \hat{{\bf N}}_i \cdot \hat{{\bf I}}_i \pm \mathrm{sgn}(\hat{{\bf N}}_i \cdot \hat{{\bf I}}_i) \sqrt{ 1-\mu_i^2 \left[  1- \left( \hat{{\bf N}}_i \cdot \hat{{\bf I}}_i\right)^2  \right] },
\end{equation}
with $\mathrm{sgn}(\hat{{\bf N}}_i \cdot \hat{{\bf I}}_i) = \hat{{\bf N}}_i \cdot \hat{{\bf I}}_i/|\hat{{\bf N}}_i \cdot \hat{{\bf I}}_i|$ for $\hat{{\bf N}}_i \cdot \hat{{\bf I}}_i \neq 0$ accounts for the relative orientation between the normal and the incident ray. For normal incidence, Eq.~(\ref{refracted_ray}) must reduce to $\hat{\mathbf{R}}_i = \hat{\mathbf{I}}_i$. Under these conditions, substituting Eq.~(\ref{gamma}) into Eq.~(\ref{refracted_ray}) confirms that the positive sign in Eq.~(\ref{gamma}) yields the physically correct root.

The formulation presented above constitutes the core of the VRT method. However, for VRT to constitute a complete framework for ray propagation in GRIN media, five additional vectorial formulations must also be considered.

\begin{figure}[t]
\centering
\includegraphics[width=.35\linewidth]{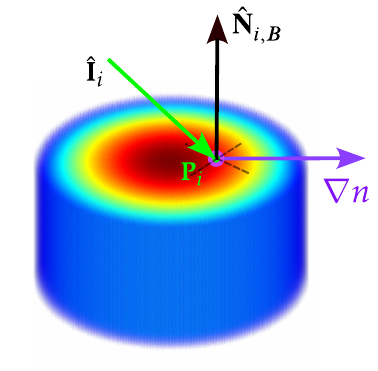}
\caption{For a point $\mathbf{P}_i$ on the planar boundary surface of a cylinder, the unit normal $\hat{\mathbf{N}}_{i}$ does not coincide with the local refractive index gradient $\nabla n$}
\label{fig:Normal_Interface}
\end{figure}

\textit{(i) Boundary normal.} VRT is not restricted to ray propagation within a freeform GRIN medium; rays may originate in an external medium and enter the GRIN domain. In this case, the boundary surface generally differs from the internal isoindicial surfaces. Consequently, when a ray crosses the boundary, the normal vector used in the vectorial Snell's law at the intersection point $\mathbf{P}_i$ is determined by the geometry of the enclosing surface, denoted as $\hat{\mathbf{N}}_{i,B}$, rather than the local refractive-index gradient (Fig.~\ref{fig:Normal_Interface}). Therefore, the next position $\mathbf{P}_i$ is obtained via Eqs.~\eqref{ray_evolution} and \eqref{Eq:Seven} by substituting $\mathbf{N}_i$ with $\hat{\mathbf{N}}_{i,B}$.

\begin{figure}[h]
\centering
\includegraphics[width=.35\linewidth]{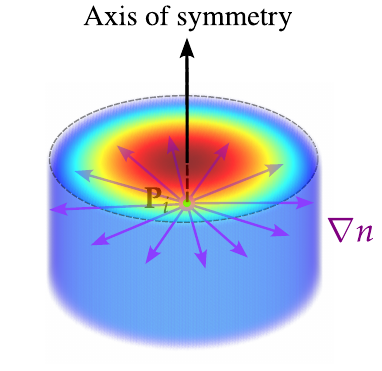}
\caption{ The normal at a point $\mathbf{P}_i$ on the axis of symmetry within the cylindrical GRIN distribution is multivalued.The figure illustrates a point located on the internal boundary of the GRIN medium along this axis.}
\label{fig:multivalued_normal}
\end{figure}

\textit{(ii) Undefined normal.} Certain refractive-index distributions may contain singular points at which the gradient $\nabla n$ is not uniquely defined. Examples include the center of spherical isoindicial surfaces or the symmetry axis of cylindrical distributions, as illustrated in Fig.~\ref{fig:multivalued_normal}. At such locations, the local normal direction cannot be uniquely determined from the refractive-index gradient. In the event that a ray reaches one of these points, propagation is continued along the current ray direction according to
\begin{eqnarray}
    \mathbf{P}_{i+1} = \mathbf{P}_i + \Delta_{i} \hat{\mathbf{I}}_i.
\end{eqnarray}

\textit{(iii) Homogeneous media.} In a homogeneous medium with refractive index $n_{\mathrm{hom}}$, the refractive-index gradient vanishes, i.e., $\nabla n = 0$, and therefore the normal vector $\hat{\mathbf{N}}_i$ cannot be obtained from the gradient. Under these conditions, the ray propagates in a straight line and the next point is determined by
\begin{equation}
    \mathbf{P}_{i+1} = \mathbf{P}_i + \Delta_{\mathrm{hom}} \hat{\mathbf{I}}_i,
\end{equation}
where $\Delta_{\mathrm{hom}} = \Delta L_0 / n_{\mathrm{hom}}$. 

\begin{figure}[h!]
\centering
\includegraphics[width=0.45\linewidth]{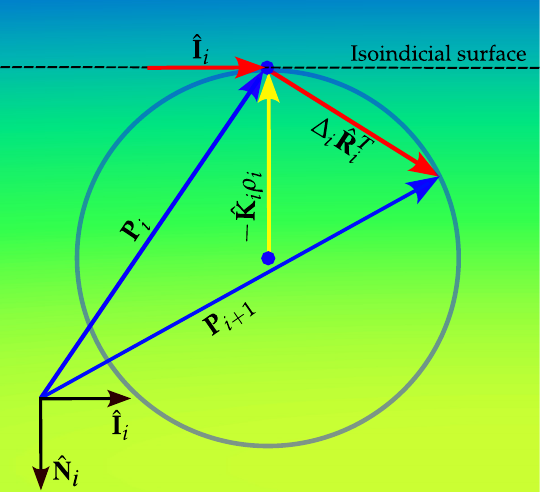}
\caption{Calculation of the subsequent point $\mathbf{P}_{i+1}$ from the current position $\mathbf{P}_i$ when the incident ray direction $\hat{\mathbf{I}}_i$ is tangent to an isoindicial surface (dashed black line) with normal vector $\hat{\mathbf{N}}_i$. }
\label{fig:Curvature_approximation}
\end{figure}

\textit{(iv) Tangential incidence.} When $\hat{\mathbf{I}}_i$ and $\hat{\mathbf{N}}_i$ are orthogonal within a GRIN medium (that is, the ray is locally tangent to an isoindicial surface) the refractive-index gradient produces a local bending of the trajectory toward regions of higher refractive index; this effect cannot be captured by the conventional Snell's law. To establish a vectorial geometrical optics rule capable of capturing this behavior, the ray equation can be rewritten under these constraints, following Stavroudis \cite{Stavroudis1972}, as
\begin{equation}
\label{solved_tangent}
     \mathbf{K} = \frac{\nabla n}{n} \quad \text{(valid for tangent incidence)},
\end{equation}
where the vector $\mathbf{K}$ represents the curvature of the ray trajectory, defined as
\begin{equation}
    \mathbf{K} = \frac{d}{ds} \left(  \frac{ d \mathbf{r}}{ds} \right).
\end{equation}
where $s$ is the arc-length parameter. To incorporate this formulation into our VRT scheme, we assume that the ray follows a locally circular path of radius $\rho_i = 1/K_i$ within the plane spanned by $\hat{\mathbf{I}}_i$ and $\hat{\mathbf{N}}_i$ (as shown in Fig.~\ref{fig:Curvature_approximation}). Under this geometric approximation, the subsequent point along the trajectory is computed as 
\begin{equation}
\mathbf{P}_{i+1} = \mathbf{P}_{i} +\Delta_i \hat{\mathbf{R}}_{i}^{T}, 
\end{equation}
where
\begin{eqnarray}
\label{P_ip1_tangent}
\hat{\mathbf{R}}_{i}^{T} =\sqrt{1-\left( \frac{\Delta_i K_i}{2} \right)^2} \hat{\mathbf{I}}_i + \frac{\Delta_i}{2} \mathbf{K}_i .
\end{eqnarray}
If the curvature vector $\mathbf{K}_i$ is approximated by
\begin{eqnarray}
\label{curvature}
\mathbf{K}_i = \frac{1}{n_i} \left(\frac{\Delta n_i}{\Delta_i} \right)\hat{\mathbf{N}}_i,\quad
\frac{\Delta n_i}{\Delta_i}
=
\frac{n(\mathbf{P}_i+\Delta_i \hat{\mathbf{N}}_i)-n(\mathbf{P}_i)}
{\Delta_i},
\end{eqnarray}
 then the expression to calculate the subsequent point simplifies to:
\begin{eqnarray}
 \mathbf{P}_{i+1} = \mathbf{P}_{i} +\Delta_i \left( \sqrt{1-\left( \frac{\Delta n_i}{2 n_i} \right)^2} \hat{\mathbf{I}}_i +  \frac{\Delta n_i}{2 n_i} \hat{\mathbf{N}}_i \right).
\end{eqnarray}


\textit{(v) Total internal reflection (TIR).} When $n_{i-1}>n_i$, the argument of the square root in Eq.~(\ref{gamma}) become negative. This condition corresponds to the occurrence of TIR within the GRIN medium when the angle of incidence exceeds the critical angle $\theta_c=\arcsin(n_i/n_{i-1})$, measured with respect to the local normal direction. Therefore, if
\begin{eqnarray}
\left(\frac{n_{i-1}}{n_i}\right)^2
\left[1-\left(\hat{\mathbf{N}}_i\cdot\hat{\mathbf{I}}_i\right)^2\right]
\geq 1,
\end{eqnarray}
the reflected ray is obtained from
\begin{eqnarray}
\mathbf{P}_{i+1}
=
\mathbf{P}_i
+
\Delta_{i-1}
\left(
\frac{
\hat{\mathbf{I}}_i
-
2\left(\hat{\mathbf{I}}_i\cdot\hat{\mathbf{N}}_i\right)\hat{\mathbf{N}}_i
}
{
\|\hat{\mathbf{I}}_i
-
2\left(\hat{\mathbf{I}}_i\cdot\hat{\mathbf{N}}_i\right)\hat{\mathbf{N}}_i\|
}
\right),
\end{eqnarray}
where $\Delta_{i-1}=\Delta L_0/n_{i-1}$.
\begin{figure*}[h]
\centering
\includegraphics[width=\linewidth]{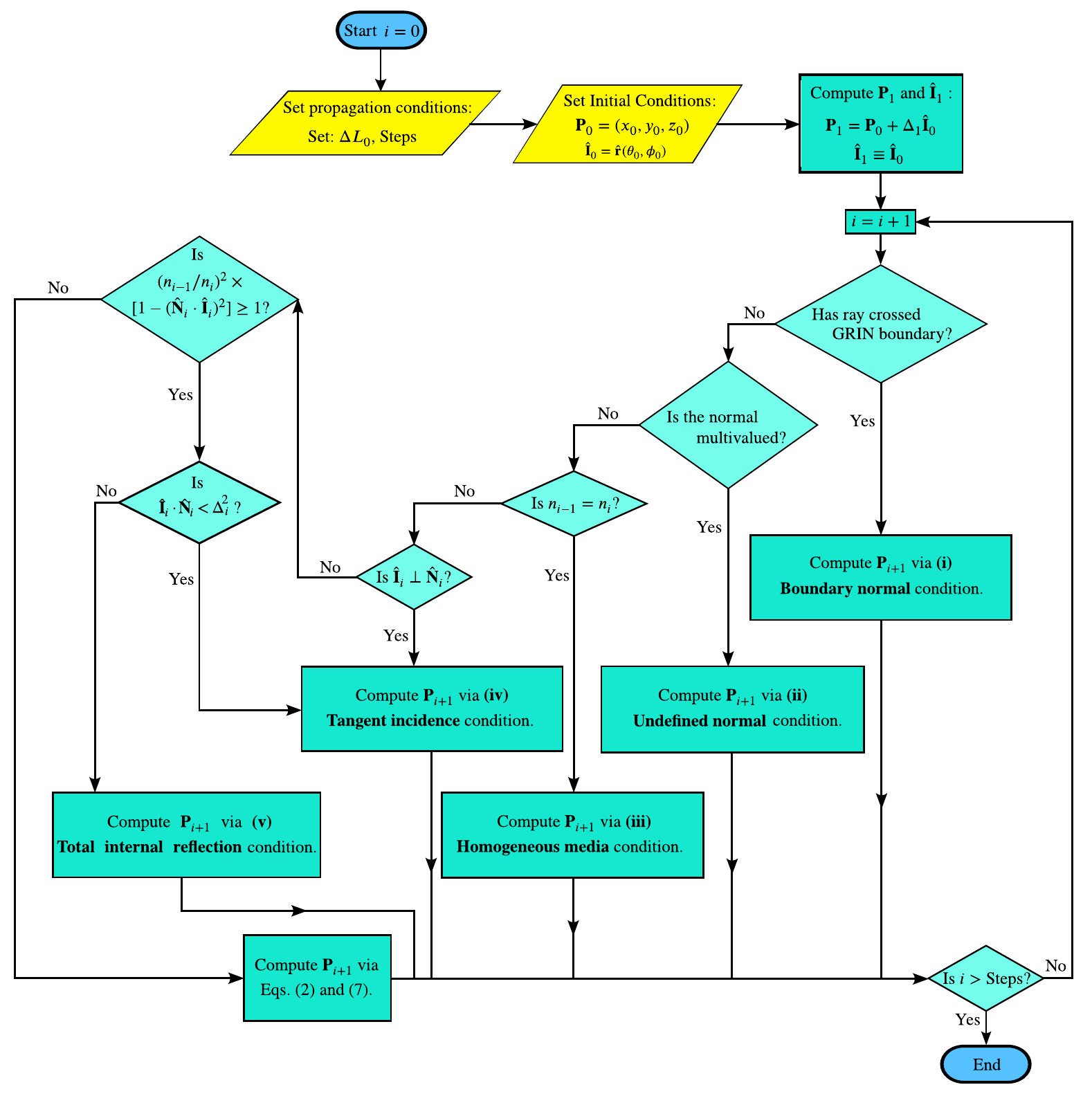}
\caption{Flowchart of the VRT algorithm. The initial direction of the ray is defined by $\hat{\mathbf{I}}_0 = \hat{\mathbf{r}}(\theta_0,\phi_0)$, where $\hat{\mathbf{r}} = (\sin\theta_0\cos\phi_0, \sin\theta_0\sin\phi_0, \cos\theta_0)$.}
\label{fig:GVRT_Flowchart}
\end{figure*}

Together with the five special cases described above, VRT provides a complete formulation for ray propagation in freeform GRIN media. The Flowchart of the VRT algorithm shown in Fig.~\ref{fig:GVRT_Flowchart} summarizes the complete propagation procedure, including interface refraction, multivalued normals, homogeneous regions, tangential incidence, and total internal reflection.

\section{VRT vs.\ Scalar Ray Tracing}

A two-dimensional scalar version of the present approach was previously reported in \cite{Chavez-Islas:25} and termed Scalar Ray Tracing (SRT). In that formulation, ray propagation is described through successive local applications of Snell’s law using scalar angular quantities. Consequently, the incident and refracted angles must be explicitly evaluated at each propagation step, requiring multiple quadrant-dependent conditional statements and increasing the algorithmic complexity.

The numerical behavior of SRT was analyzed in \cite{Chavez-Islas:26} using a two-dimensional Luneburg lens. The results showed that the root-mean-square error decreases approximately linearly with the OPL step size, indicating first-order convergence analogous to Euler's method (RK1). Since VRT is based on the same propagation principle and employs the same constant-OPL stepping scheme, an equivalent convergence behavior is expected when restricted to two-dimensional configurations.

The primary motivation for VRT is therefore not to alter the convergence properties of the method, but rather to provide a natural extension to arbitrary three-dimensional geometries. While SRT is well suited for planar configurations, its extension to general freeform three-dimensional media would require the explicit treatment of non-planar trajectories and additional angular conventions. In contrast, VRT operates directly with vector quantities, allowing the same formulation to be applied in both two- and three-dimensional environments.

Furthermore, VRT was developed to account for all propagation scenarios that may arise in freeform GRIN media immersed in homogeneous environments. These include interface refraction, continuous gradient-induced ray deviations, total internal reflection, homogeneous regions, and singular points where the refractive-index gradient is not uniquely defined. As summarized in the flowchart of the VRT algorithm (see Fig. \ref{fig:GVRT_Flowchart}), all these situations are incorporated within a single propagation framework.

An additional advantage of VRT is its computational efficiency. Because the method avoids the explicit evaluation of angular quantities and the associated quadrant-dependent conditions required by SRT, the resulting implementation is considerably simpler. For a single ray propagating through a Luneburg lens using an OPL step size of $10^{-4}$ and 26,000 propagation steps, VRT is 1.73 times faster than SRT. For a bundle of 100 rays propagated under the same conditions, VRT remains 1.60 times faster than SRT.

Therefore, VRT may be viewed as the natural vectorial generalization of SRT, preserving its local refraction-based propagation scheme while extending its applicability to arbitrary three-dimensional freeform GRIN media and incorporating the physical propagation scenarios required for practical optical modeling and design.

\section{Error analysis}
\label{sec:errores}
 
Since the primary motivation of VRT is optical analysis and design in freeform gradient-index media, its accuracy is evaluated using a fully three-dimensional benchmark for which an analytical solution exists. This enables a direct comparison between the VRT and analytical ray trajectories under genuinely three-dimensional propagation conditions. Specifically, we consider a cylindrical GRIN medium with a parabolic-index fiber distribution \cite{Lakshminarayanan,Gomez-Correa:21}, whose refractive-index profile is given by
\begin{equation} \label{eq:parabolic_n}
    n^2(\rho)=\begin{cases}
       n_c^2\left(1-2\Delta\frac{\rho^2}{R^2}\right)\quad\text{for}&0<\rho \leq R \\
       n_c^2\left(1-2\Delta\right)\quad\text{for}&\rho>R
    \end{cases},
\end{equation}
where $n_c$ corresponds to the constant refractive index value at $\rho=0$, with $\rho=\sqrt{x^2+y^2}$, and $R$ represents the radius of the fiber core. Note that the interface between the core and the cladding (external constant medium) is located at $\rho=R$.

\begin{figure*}[h]
\centering
\includegraphics[width=0.7\linewidth]{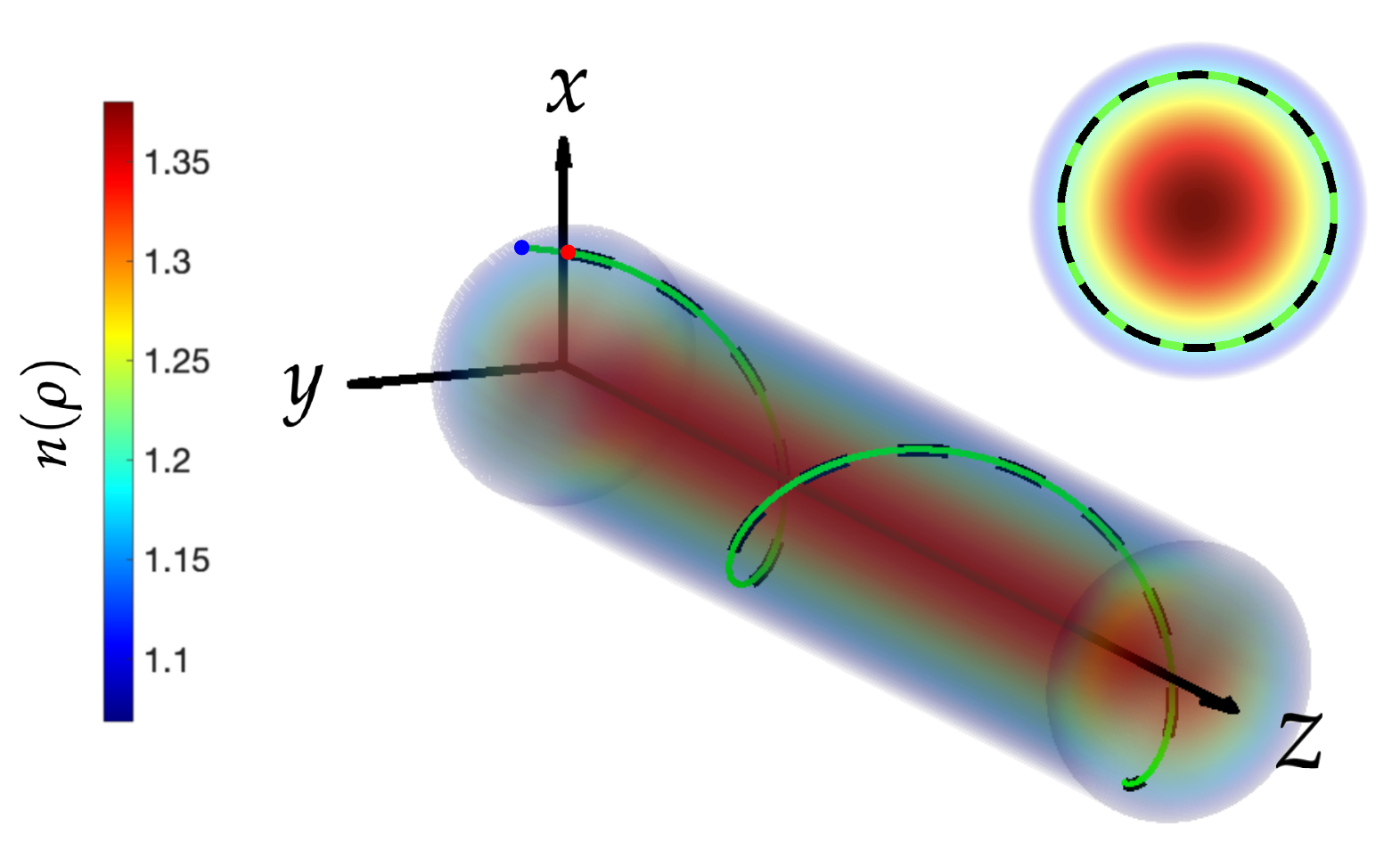}
\caption{Comparison between the VRT and analytical solutions for a cylindrical GRIN medium with a parabolic-index fiber distribution. The projection of the ray trajectory onto the $xy$-plane shows a helical path at a constant radial distance $\rho = 4$. The ray originates at $\mathbf{P}_0 = (4, -2.6128, -3.0288)$ (blue point) and is directed to impact the base of the fiber ($z = 0$) at the point 
$\mathbf{P}_I = (4, 0, 0)$ (red point).}
\label{fig:2}
\end{figure*}

In Fig.~\ref{fig:2}, we present the analytical (black dashed line) and the corresponding numerical (solid green line) solutions for a ray that initially lies in the $yz$-plane, at $x = 4$, and impacts the base of the cylinder, $z = 0$, at an angle of $40.78^{\circ}$ relative to the $z$-axis, giving rise to a helical trajectory within the GRIN medium. For the numerical computation, a fixed OPL increment of $\Delta L_0 = 1 \times 10^{-4}$ was chosen. The radius of the GRIN cylinder was set to $R=5$, with $n_c=1.38$ and $\Delta=0.2$ (all spatial parameters are given in arbitrary units). Additionally, for illustration purposes, Fig.~\ref{fig:2} also shows the transverse projection, normal to the propagation axis, where it can be seen that the ray propagates at a constant $\rho$, following the expected helical trajectory \cite{Gomez-Correa:21,Lakshminarayanan}. Visually, we observe a very good match in both images.

To quantify the agreement between the VRT and analytical solutions, a three-dimensional trajectory benchmark was performed using the transverse root mean square error (RMSE). Since both trajectories are compared at the same axial positions $z_i$, the error is evaluated in the transverse coordinates as
\begin{equation} \label{eq:rmse}
\text{RMSE} = \sqrt{\frac{1}{m}\sum_{i=1}^m \left( \left[x(z_i) - x_i(z_i)\right]^2 + \left[y(z_i) - y_i(z_i)\right]^2 \right)},
\end{equation}
where $x(z_i)$ and $y(z_i)$ correspond to the analytical solution evaluated at the axial position $z_i$, while $x_i$ and $y_i$ denote the coordinates obtained from the VRT solution evaluated at the same axial position. Here, $m$ indicates the total number of propagation points. Using a constant OPL step size of $\Delta L_0 = 1\times10^{-4}$ and $m = 808,360$ propagation points, the RMSE value was $5.0892\times10^{-5}$. The computational time for both ray trajectory computation and rendering was 0.5157 seconds and was obtained using MATLAB R2024b on a MacBook Pro equipped with an Apple M4 chip and 24 GB of RAM.

\section{Ray tracing in freeform three-dimensional GRIN media}

Having established the accuracy and numerical consistency of the VRT framework through comparison with analytical solutions, we now consider two representative freeform configurations corresponding to fully three-dimensional, non-symmetric GRIN media, where no analytical description of the ray trajectories is available.

The first configuration corresponds to a corrective freeform cylindrical GRIN lens proposed by T. Yang et al.~\cite{Yang:20}. In this case, the refractive-index distribution is defined as
\begin{eqnarray}
n(x,y,z) = n_0 + \Delta n \sum_{m,n} c_{m,n}\, Z_m\left(x'(x),y'(y)\right)\, V_n\left(z'(z)\right),
\end{eqnarray}
where the transverse dependence is described by the Fringe Zernike polynomials [$Z_m(\cdot,\cdot)$] ~\cite{Yang:20,Niu_2022} and the axial variation by Legendre polynomials [$V_n(\cdot)$]. The normalized coordinates are defined as
\begin{eqnarray}
x'(x) = \frac{x}{\rho_{\max}}, \qquad
y'(y) = \frac{y}{\rho_{\max}}, \qquad
z'(z) = \frac{2z}{z_{\max}} - 1,
\end{eqnarray}
where $\rho_{\max}$ and $z_{\max}$ denote the radius and thickness of the cylindrical boundary, respectively, satisfying $x'^2+y'^2\leq 1$ and $-1\leq z' \leq 1$. Furthermore,
\begin{eqnarray}
n_0 = \frac{n_{\mathrm{hi}} + n_{\mathrm{lo}}}{2},
\qquad
\Delta n = \frac{n_{\mathrm{hi}} - n_{\mathrm{lo}}}{2}.
\end{eqnarray}
This approach is known as the binary linear GRIN composition model, where $n_{\mathrm{hi}}$ and $n_{\mathrm{lo}}$ are the refractive indices of two homogeneous parent materials. Such a corrector lens model was fabricated and reported in Ref.~\cite{Yang:20} with $n_{\mathrm{lo}} = 1.5148$ and $n_{\mathrm{hi}} = 1.5162$. To establish a direct comparison, we adopt their spatial parameters but retain them as arbitrary units. Accordingly, the lens radius $\rho_{\text{max}}$ and thickness $z_{\text{max}}$ are set to 4.2 and 2.5 arbitrary units, respectively. The corresponding nonzero expansion coefficients are $c_{5,0} = -0.2368$, $c_{8,0} = 0.0249$, $c_{9,0} = 0.0011$, $c_{11,0} = -0.0078$, $c_{12,0} = -0.0013$, $c_{8,1} = 0.5548$, and $c_{11,1} = -0.1731$.

\begin{figure*}[h]
\centering
\includegraphics[width=0.8\linewidth]{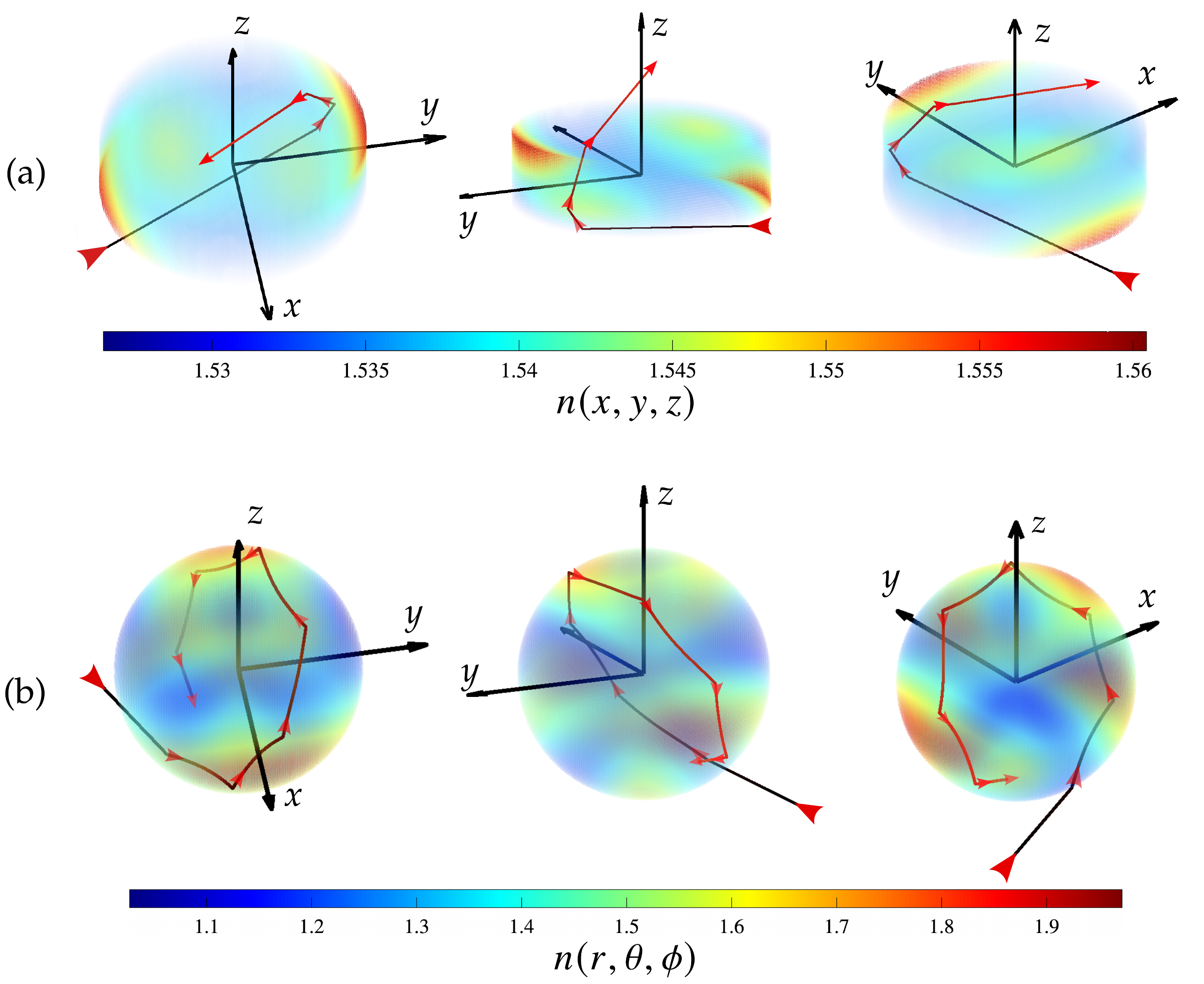}
\caption{(a) VRT trajectory starting from $P_0^1 = (1,-5,-1.5)$ with initial direction in spherical coordinates  $\theta_01 = 80^\circ$ and $\phi_0^1 = 110^\circ$ for a corrective freeform cylindrical lens. 
(b) VRT in a polynomial GRIN medium with $(r,\theta,\phi)$ dependence, starting from $P_0^1 = (-1,-1,-1)$ with $\theta_{0}^{1} = 80^\circ$ and $\phi_0^1 = 20^\circ$. The large red arrow indicates the starting point of the ray, while the subsequent arrow marks its entry point into the lens. All following red arrows represent TIR events at the boundary of the GRIN medium. (a) The penultimate arrow indicates the location where the ray exits the lens.
 }
\label{fig:3}
\end{figure*}

An example of a ray trajectory computed with VRT is shown in Fig.~\ref{fig:3}(a). Due to the relatively small refractive-index variation ($n_{\mathrm{lo}}=1.5261$, $n_{\mathrm{hi}}=1.5608$), the ray undergoes a smooth and moderate deviation inside the lens. The transition from the external homogeneous region into the GRIN medium is consistently captured, and the subsequent evolution follows the expected gradual bending induced by the index distribution.

The second configuration considers a spherical freeform GRIN medium with refractive-index distribution defined in spherical coordinates $(r,\theta,\phi)$, 
\begin{eqnarray}
n(r,\theta,\phi) = a_0 + \sum_{m=1}^{3} a_m\, G_m(r,\theta,\phi),
\end{eqnarray}
where $G_1 = 0.3\,(r^2 - 1)$, $G_2 = r\cos(5\theta)$, and $G_3 = r^2 \sin^2\theta \cos(2\phi)$ represents a set of customized functions, with $a_0 = 1.5$, $a_1 = 1$, $a_2 = -0.2$, and $a_3 = 0.3$. Here, the radial coordinate is dimensionless, having been normalized such that the maximum radius of the sphere is defined as $r_{\text{max}} = 1$.

The resulting trajectory is shown in Fig.~\ref{fig:3}(b). Observe that the freeform gradient index distribution leads to pronounced curvature and complex directional changes of the propagating ray. Here, we remark that the VRT method handles in a natural way boundary interactions, including total internal reflection, without requiring additional modeling assumptions.

In both configurations, the trajectory is fully determined by its respective initial conditions, namely the initial point $\mathbf{P}_0^{1,2}=(x_0^{1,2},y_0^{1,2},z_0^{1,2})$ and the initial direction $\hat{\mathbf{I}}_0^{1,2} = (\sin\theta_0^{1,2}\cos\phi_0^{1,2}, \sin\theta_0^{1,2}\sin\phi_0^{1,2}, \cos\theta_0^{1,2})$, where the superscripts $1$ and $2$ denote the corrective freeform cylindrical GRIN lens and the spherical freeform GRIN medium, respectively. From these initial conditions, the VRT method constructs the ray path iteratively, remaining applicable in arbitrary refractive-index distributions without imposing symmetry constraints or relying on predefined analytical forms.

For completeness, Fig.~\ref{fig:3}(a) and Fig.~\ref{fig:3}(b) also present different views of the trajectory alongside its three-dimensional representation. The color gradient along the ray trajectory, ranging from black to red, indicates the direction of propagation from the launch point to the exit point.

\section{Optical design using VRT}

One of the central requirements of any ray tracing framework is its ability to support optical design. In this section, we demonstrate that the VRT formulation can be directly employed for this purpose by constructing two gradient-index lenses with prescribed focusing properties.

As a first design example, we consider the design of a cylindrical GRIN lens with a prescribed focal distance and strongly reduced longitudinal spherical aberration (LSA). The objective is to determine a refractive-index distribution that causes rays launched at different heights to converge as closely as possible to the same focal point, thereby approaching a quasi-stigmatic focusing condition.

To this end, we consider a cylindrical GRIN lens described by a two-parameter refractive-index distribution, denoted here as cylindrical N2P-GRIN lens. The refractive-index distribution is defined as
\begin{equation}
\label{GRIN_cylinder}
    n_{\beta, \gamma}(\rho) = 
    \begin{cases} 
        \left[ 1 + \left( \frac{R^2 - (\rho/2)^2}{F^2} \right)^{\beta} \right]^{\gamma} & \text{if } z \in [-L, 0] \text{ and } 0 \leq \rho \leq R \\
        1 & \text{otherwise},
    \end{cases}
\end{equation}
where $R$ is the cylinder radius, $F$ is the focal distance, $L$ is the lens thickness, and $\beta$ and $\gamma$ are free parameters to be determined.

The performance of the lens is evaluated in terms of the LSA, quantified through the root mean square error (RMSE),
\begin{equation}
\text{RMSE}_{\mathrm{LSA}} = \sqrt{\frac{1}{N}\sum_{i=1}^{N}\text{LSA}_{i}^{2}},
\label{Eq:Err}
\end{equation}
where
\begin{equation}
\text{LSA}_i = F - z_{\mathrm{Int},i},
\end{equation}
and $z_{\mathrm{Int},i}$ denotes the axial coordinate at which the $i$-th ray intersects the optical axis.

\begin{figure}[h]
\centering
\includegraphics[width=0.5\linewidth]{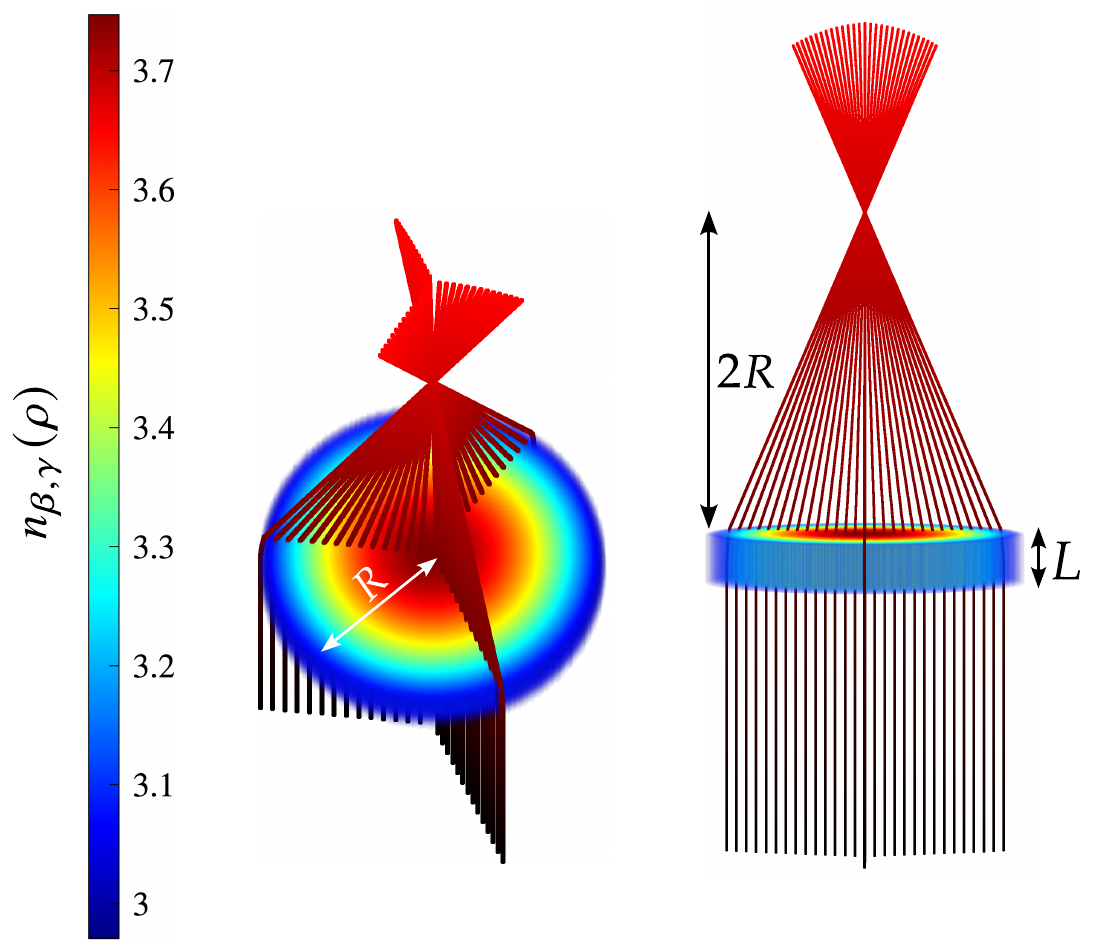}
\caption{VRT through a radial GRIN medium, illustrating light focusing from two different perspectives. The incident rays lie in three semiplanes sharing the optical axis as a common boundary; these planes are equiangularly separated.}
\label{fig:optimized_cylinder}
\end{figure}

Due to the cylindrical symmetry of the system, the optimization process is restricted to the $yz$-plane. A set of $N=10$ rays is considered, uniformly distributed along the $y$-axis, excluding $y=0$. The maximum ray height is set to $0.995R$, and the propagation is performed using a constant OPL step $\Delta L_0 = 1 \times 10^{-4} R$.

The optimization of $\beta$ and $\gamma$ is carried out through a successive refinement strategy. Starting from an initial parameter range, a discrete set of values is sampled and the minimum of $\text{RMSE}_{\mathrm{LSA}}$ is identified. The search interval is then iteratively reduced around this minimum until convergence is achieved.

\begin{figure}[h]
\centering
\includegraphics[width=0.5\linewidth]{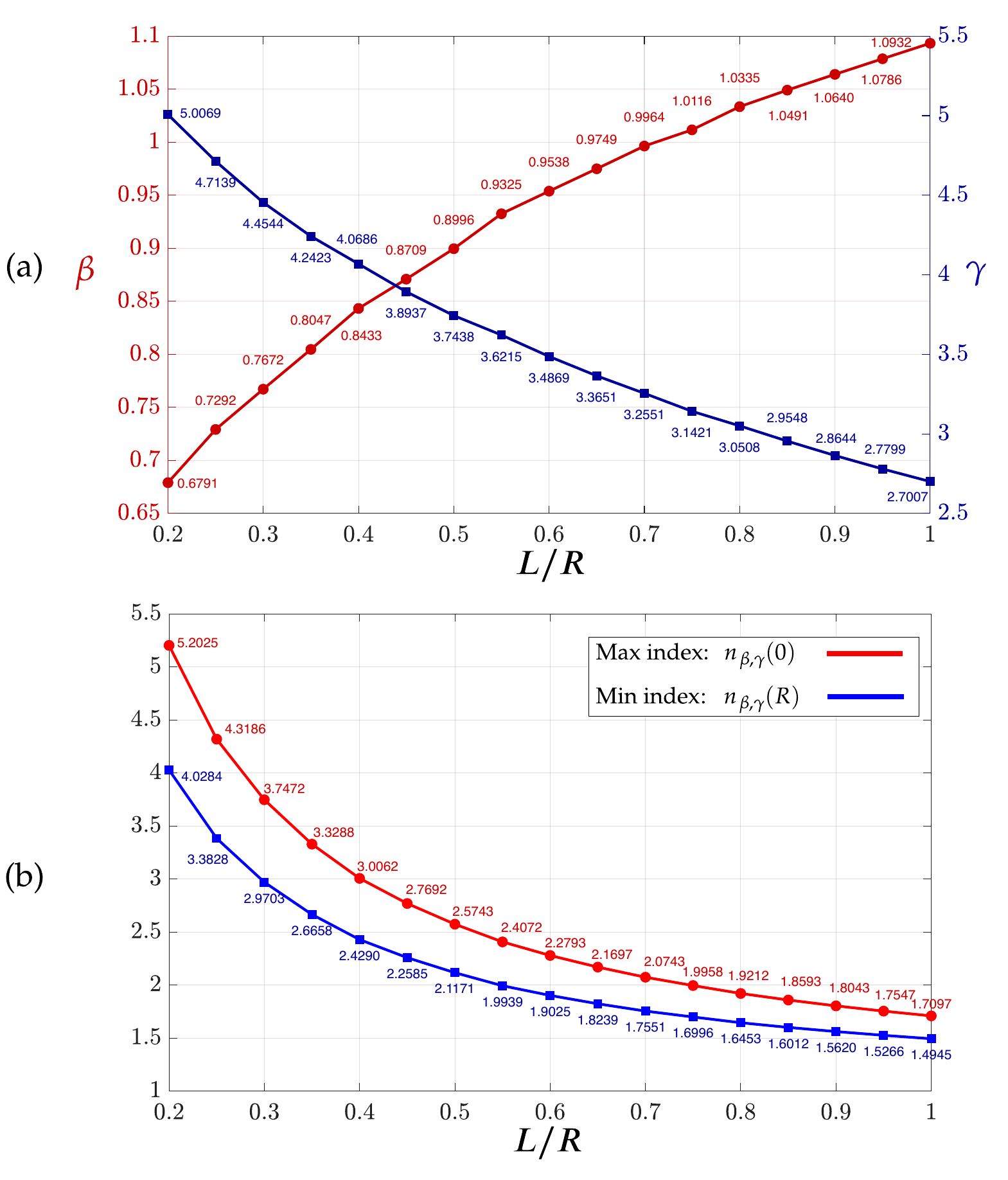}
\caption{Figure~(a) shows the dependence of parameters $\beta$ and $\gamma$ as a function of the cylinder width $L$ for the focusing lens, while Figure~(b) displays the corresponding maximum and minimum refractive index values.  }
\label{fig:L_vs}
\end{figure}

The optimized configuration for a cylindrical lens with thickness $L=0.3R$ and focal distance $F=2R$ is shown in Fig.~\ref{fig:optimized_cylinder}. The dependence of the optimized parameters on the lens thickness is shown in Fig.~\ref{fig:L_vs}(a) for the same focal distance. As $L$ varies from $0.2R$ to $R$, the parameter $\gamma$ exhibits a dominant influence on the focusing behavior, increasing from $2.7007$ to $5.0069$, while $\beta$ varies more moderately from $0.6791$ to $1.0932$. The corresponding $\text{RMSE}_{\mathrm{LSA}}$ remains on the order of $(2.4014 \pm 0.2504) \times 10^{-3}R$.

A numerical exploration indicates that as the lens thickness decreases below $0.2R$, total internal reflection hinders the optimization process. 

Finally, Fig.~\ref{fig:L_vs}(b) shows the corresponding variation of the maximum and minimum refractive index values. As the lens thickness decreases, the required refractive index contrast increases, reflecting the reduced spatial extent available to redirect the rays toward the focal point.

As a second design example, we consider a freeform GRIN lens designed to generate multiple focal regions from parallel incident ray bundles. In contrast to the previous cylindrical N2P-GRIN lens, where the objective was to minimize the LSA around a prescribed focal point, this example illustrates how VRT can be used to analyze and design a more complex optical element capable of redistributing incoming light into several spatially separated foci.

\begin{figure}[h]
\centering
\includegraphics[width=0.5\linewidth]{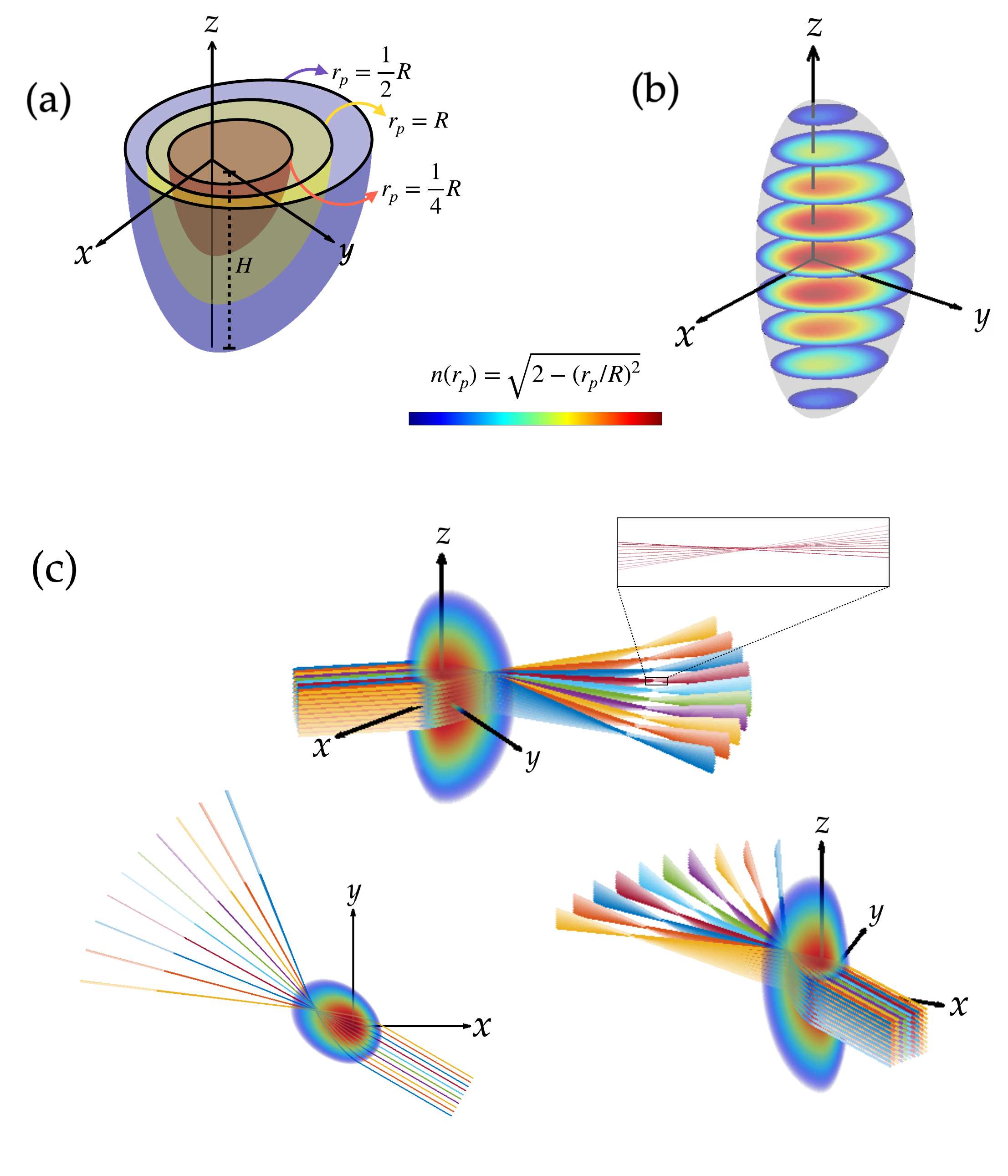}
\caption{(a) Half-dome structure illustrating the construction of the three-dimensional freeform GRIN medium. (b) Slices illustrating the three-dimensional freeform geometry generated from Eq.~\eqref{eq:freeform} with $a = 0.35$, $b = 0.15$, and $m = 2$. (c) Ray propagation through the freeform GRIN medium obtained using the VRT method. Different colors denote distinct columns of incident rays to highlight trajectories and the formation of focal regions.}
\label{fig:freeform}
\end{figure}

The freeform geometry is generated from an arbitrary polar function on the $x$-$y$ plane of the form \begin{equation} \label{eq:freeform}
\rho(\theta;r_p)=r_p\frac{a}{1+a\cos(\theta)+b\sin(m\theta)},
\end{equation}
where $\theta$ is the polar angle, $\rho$ is the radial coordinate, $a$, $b$ and $m$ are real dimensionless constants and $r_p$ is a scaling parameter expressed in arbitrary spatial units that determines the size of the resulting parametric polar curve, taking real values in the interval $[0,R]$. Each value of $r_p$ within this interval defines an isoindicial curve, as shown on the $x$-$y$ plane in Fig.~\ref{fig:freeform}(a). The planar geometry enclosed by each isoindicial curve is then extended into three dimensions by introducing an additional parameter that expands the profile along both directions of the axis perpendicular to the plane ($z$-axis) up to a maximum height $H$, creating dome-like surfaces bounded by the parametric curve defined in Eq.~\eqref{eq:freeform}. In this manner, each isoindicial curve described by Eq.~\eqref{eq:freeform} evolves into a closed shell that now represents an isoindicial surface, as illustrated in Fig.~\ref{fig:freeform}(a), where a cross-sectional view of the lens reveals the symmetric dome-like structure. Consequently, a GRIN distribution can be constructed by mapping the parameter $r_p$ to refractive index values according to a prescribed distribution $n\left(r_p(x,y,z)\right)$~\cite{ThesisVrani}.

By selecting $a=0.35$, $b=0.15$, and $m=2$ in Eq.\eqref{eq:freeform}, and adopting the refractive index distribution of the Luneburg lens $n(r_p) = \sqrt{2-\left(r_p/R\right)^2}$, where $R$ denotes the maximum lens radius \cite{Luneburg}, the resulting three-dimensional freeform geometry is obtained, as presented via sequential cross-sectional slices in Fig.~\ref{fig:freeform}(b). The implementation of the VRT method within the associated GRIN medium is shown in Fig.\ref{fig:freeform}(c).
For visualization purposes, a distinct color and its corresponding shades were assigned to each parallel column of rays entering the GRIN medium along the positive $y$-direction. This facilitates the observation of how each ray bundle bends while propagating through the medium and how, upon exiting it, a focal region is formed for each group of rays.
The propagation behavior of rays within this freeform medium may offer potential advantages for incident light multiplexing applications \cite{paperAnel}. Although the present example primarily serves to demonstrate the applicability of the VRT method to arbitrary optical geometries, more intricate structures may be designed to redirect light in a prescribed manner, thereby enabling the development of customized lens systems.

\section{Conclusions}

In this work, we introduced the Vectorial Ray Tracing (VRT) method, a ray tracing framework in which propagation is modeled through successive local applications of vectorial relations of geometrical optics. Combined with a constant optical path length stepping scheme, the formulation enables the propagation of rays in homogeneous media, discretized refractive-index distributions, and continuous three-dimensional GRIN media within a unified framework.

The method was validated against an analytical solution using a fully three-dimensional benchmark, showing excellent agreement between the analytical and numerical trajectories. In addition, VRT naturally incorporates vectorial rules of geometrical optics to manage refraction at the boundary of the GRIN medium, regions where the refractive-index gradient is not uniquely defined, homogeneous regions, local bending of the trajectory when the ray is tangent to the iso-surface, and total internal reflection. The resulting formulation is straightforward to implement and applicable to arbitrary freeform GRIN geometries.

Two design examples were presented to illustrate the capabilities of the method. The first demonstrated the design of a cylindrical GRIN lens with strongly reduced longitudinal spherical aberration, while the second showed the construction of a freeform GRIN structure capable of generating multiple focal regions. These examples highlight the applicability of VRT not only for ray propagation analysis, but also for the design and optimization of GRIN-based optical systems.

The proposed framework provides a general approach for studying light propagation in complex three-dimensional inhomogeneous media and may be extended to a broad range of optical design problems involving arbitrary refractive-index distributions.

\begin{acknowledgments}
The first two authors thank the Secretar\'{i}a de Ciencia, Humanidades, Tecnolog\'{i}a e Innovaci\'{o}n (SECIHTI) for the support provided through the awarded scholarships (VCI: 1348421; EER: 640385). FHMV also acknowledges support from SECIHTI through project IIxM 7309 (CIR/072/2024). 
\end{acknowledgments}


\bibliography{OLTRef}   
\bibliographystyle{spiejour}   

\end{document}